\documentclass[pra,twocolumn,tightenlines,showpacs,nofootinbib]{revtex4}
\usepackage{multirow}
\usepackage{bm,dcolumn,amsmath,graphicx}
\begin{document}

\title{Polarizabilities, Stark shifts, and lifetimes of In atom}
\author{M.~S.~Safronova$^{1,2}$}
\author{U.~I.~Safronova$^{3,4}$}
\author{S.~G.~Porsev$^{1,5}$}

\affiliation{$^1$Department of Physics and Astronomy, University of Delaware,
Newark, Delaware 19716, USA\\
$^2$Joint Quantum Institute, National Institute of Standards and Technology and \\
the University of Maryland, Gaithersburg, Maryland, 20899-8410, USA\\
$^3$Physics Department, University of Nevada, Reno, Nevada 89557,
\\$^4$Department of Physics,  University of Notre Dame, Notre Dame, IN 46556 \\
$^5$Petersburg Nuclear Physics Institute, Gatchina, Leningrad District, 188300, Russia }
\begin{abstract}
We evaluate the polarizabilities of the $5p_{1/2}$, $6s$, $6p_{1/2}$, and $6p_{3/2}$ states of In using two different
high-precision relativistic methods: linearized coupled-cluster approach where  single, double and
partial triple excitations of the Dirac-Fock wave function are included to all orders of perturbation theory
and an approach that combines the configuration interaction  and the
coupled-cluster method. Extensive comparison of the accuracy of these methods is carried out.
The uncertainties of all recommended values are evaluated. Our result for the $6s-5p_{1/2}$
Stark shift is in excellent agreement with the recent measurement [Ranjit {\it et al.}, arXiv:1302.0821v1].
Combining our calculation with
this precision measurement allows us to infer the values of the $6p_{1/2}$ and $6p_{3/2}$ lifetimes in In with
0.8\% accuracy. Our predictions for the $6p_{3/2}$ scalar and tensor polarizabilities may be combined with the future measurement of the
$6s-6p_{3/2}$ Stark shift to accurately determine the lifetimes of the $5d_j$ states.
\end{abstract}

\date{\today}

\maketitle
\section{Introduction}

An indium atom represents an excellent system to compare the accuracy of different high-precision  theoretical methods
since it may be considered as both the monovalent system (assuming a closed $5s^2$ core) and a trivalent system. In
this work, we use both linearized coupled-cluster approach where  single, double and
partial triple excitations (LCCSDpT) of the Dirac-Fock wave function are included to all orders
and a method combining the configuration interaction (CI) and
the coupled-cluster method (CI+all-order).

The monovalent LCCSDpT method has been applied to a large number of neutral and ionized monovalent systems and yielded
very accurate predictions for a number of atomic properties (see review \cite{SafJoh08} and references therein). It has
been used for a variety of applications ranging from the study of fundamental symmetries~\cite{PorBelDer09,Saf11} to
study of the degenerate quantum gases and quantum information~\cite{SafWilCla03}. This method has been tested against
all high-precision alkali and monovalent ion experimental values which allowed to establish a systematic procedure to
evaluate its uncertainties even when no experimental data exist~\cite{SafSaf11}.

The CI+all-order method was recently developed
for the treatment of the more complicated systems \cite{SafKozJoh09}. It has been tested on a variety of divalent systems
\cite{SafPorCla12,SafKozCla11,SafKozCla12,SafKozSaf12,SafPorKoz12} and applied to
Tl \cite{PorSafKoz12}. However, there are far less experimental benchmark experiments for the properties of
the divalent and trivalent systems in comparison with the alkalis.
A recent high-precision (0.27\%) measurement of the $6s-5p_{1/2}$ Stark shift in
In~\cite{RanSchLor13} provides an excellent opportunity to compare the accuracy of the CI+all-order and
LCCSDpT approaches, and test the procedure for the evaluation of the uncertainties of theoretical data.
In turn, the combination of the Stark shift measurement  \cite{RanSchLor13} and our calculations allows us to
infer the values of the $6p_{1/2}$ and $6p_{3/2}$ lifetimes in In with
0.8\% accuracy. It is extremely difficult to achieve such high accuracy via direct lifetime measurements.
Moreover, we predict the values of the $6p_{3/2}$ scalar and tensor polarizabilities which can be soon tested against the
future measurement of the $6s-6p_{3/2}$ Stark shift \cite{RanSchLor13} and will allow to determine the $5d_{j}$
lifetimes.

Precise knowledge of In properties is of interest to the study of the fundamental symmetrises, including parity
violation and search for the permanent electric-dipole moment,  since it is similar to Tl (also group IIIB).
Understanding of the theoretical and experimental uncertainties in In can be used in Tl studies. Both EDM
\cite{RegComSch02} and parity-violation studies \cite{EdwPhiBai95,VetMeeMaj95} had been carried our in Tl. The theory
accuracy in Tl is still below the experimental accuracy, hindering further parity violation studies with this system. A
recent controversy regarding the calculated values of Tl EDM enhancement factor is discussed in detail in
\cite{PorSafKoz12}.

In-like ions are excellent candidates for a search for the variation of the fine-structure constant $\alpha$. Despite
very large ionization energies, certain ions have transitions that lie in the optical range due to level crossing  and
are very sensitive to $\alpha$-variation \cite{BerDzuFla10}. In-like ions  are particulary well suite for the
experimental search for such transitions \cite{SafDzuFla13}. In fact, In-like isoelectronic sequence has by far the
largest number of ions with  long-lived metastable states with transition frequencies between 170 and 3000 nm, high
sensitivity to $\alpha$-variation, and stable isotopes \cite{SafDzuFla13}.  One of the main obstacles for the
experimental work in this direction is the lack of any experimental data for these systems and difficulty of accurate
theoretical predictions. Testing the CI+all-order method on neutral In provides important information of the accuracy
of this approach for further studies of the In-like ions.

\section{Method}
We use two different relativistic high-precision approaches for all calculations in this work. Such comparison of
these two methods for the same system have never been done before.
The first method (LCCSDpT) has been extremely
successfully in predicting properties of alkali-metal atoms and other monovalent ions~\cite{SafJoh08}.
It can also be applied to In by treating $5s^2$ shell as a part of the
[$1s^2 2s^2 2p^6 3s^2 3p^6 3d^{10} 4s^2 4p^6 4d^{10} 5s^2$] core. The disadvantage of
this approach is its inability to explicitly treat three-particle states, such as $5s5p^2$.
However, LCCSDpT method produced the results for the $6p_{1/2}-7s$ and  $6p_{1/2}-7p_{1/2}$ Stark
shifts \cite{SafJohSaf06} of Tl, which is a similar Group IIIB system, in excellent agreement
with the experiments ~\cite{DorFriSpe02,DemBudCom94}. For convenience, we will refer to the LCCSDpT
approach as the CC method in text and tables below.

The CI+all-order method (we also refer to it as CI+All in text and tables below) was developed in \cite{SafKozJoh09}.
It allows us to treat In as a three-particle system, so all three electrons above the $4d^{10}$ shell are considered
valence. In this approach, the CC method is first used to accurately describe core-core and core-valence correlation
and to incorporate them into the effective Hamiltonian.   Therefore, the core-core and core-valence sectors of the
correlation corrections for systems with few valence electrons will be treated with the same accuracy as in the
all-order approach for monovalent atoms. Then, the CI method is used to treat valence-valence correlations. Since the
CI space includes only three electrons, it can be made essentially complete. This method allows to include correlation
corrections to the wave functions in a more complete way than the CC approach. In particular it is capable to
accurately account for  the configuration mixing.

The CI+all-order method yielded accurate wave functions for calculations of
such atomic properties as lifetimes, polarizabilities, hyperfine structure
constants, etc. for a number of divalent systems and Tl
\cite{SafKozJoh09,SafKozCla11,PorSafKoz12,SafPorKoz12,SafKozCla12,SafKozSaf12,SafPorCla12}.
However, the various types of the corrections to the effective dipole operator $D_{\rm eff}$ are included in a more
complete way in the CC approach at the present time. Therefore, both approaches are complementary
and the difference in the results can serve as an estimate of the uncertainties.
\begin{table}
\caption{Comparison of the CC and CI+all-order (labeled as ``CI+All'') energies of
In levels with experiment \cite{RalKraRea11}. Three-electron binding energies are given in the first row.
The energies in other rows are given relative to the ground
state. Corresponding relative differences of these two calculations with experiment
are given in the corresponding columns labeled ``Diff.'' in \%. The
$5s^26p~^2\!P_{1/2}^o - 5s^26s~^2\!S_{1/2}$ and $5s^25d~^2\!D_{5/2}- 5s^26p~^2\!P_{3/2}^o$
 transition energies are given in
the last row.} \label{tab1}
\begin{ruledtabular}
\begin{tabular}{llrrrrr}
\multicolumn{2}{l}{State} & \multicolumn{1}{c}{Expt.} & \multicolumn{1}{c}{CC} & \multicolumn{1}{c}{Diff.} &
\multicolumn{1}{c}{CI+All} & \multicolumn{1}{c}{Diff.} \\
\hline
$5s^2 5p $      &$^2\!P_{1/2}^o$&   425060&         &          & 425719&   0.15\%  \\
                &$^2\!P_{3/2}^o$&   2213  &  2168   &   2.02\% &  2195 &   0.82\%    \\
$5s^2 6p $      &$^2\!P_{1/2}^o$&   31817 &  31468  &   1.10\% &  31805&   0.04\%    \\
                &$^2\!P_{3/2}^o$&   32115 &  31769  &   1.08\% &  32104&   0.03\%    \\
$5s^2 7p $      &$^2\!P_{1/2}^o$&   38861 &  38513  &   0.90\% &  38911&  -0.13\%    \\
                &$^2\!P_{3/2}^o$&   38973 &  38625  &   0.89\%  & 39023&  -0.13\%   \\ [0.4pc]
$5s^2 6s $      &$^2\!S_{1/2}$  &   24373 &  23862  &   2.10\%  & 24272&   0.41\%   \\
$5s^2 5d $      &$^2\!D_{3/2}$  &   32892 &  32563  &   1.00\%  & 32836&   0.17\%   \\
                &$^2\!D_{5/2}$  &   32916 &  32754  &   0.49\%  & 32863&   0.16\%   \\
$5s 5p^2 $      &$^4\!P_{1/2}$  &   34978 &         &           & 35299&  -0.92\%   \\
                &$^4\!P_{3/2}$  &   36021 &         &           & 36346&  -0.90\%   \\
                &$^4\!P_{5/2}$  &   37452 &         &           & 37770&  -0.85\%   \\
$5s^2 7s $      &$^2\!S_{1/2}$  &   36302 &  35928  &   1.03\%  & 36284&   0.05\%    \\  [0.4pc]
\multicolumn{2}{l}{$\Delta(6p_{1/2}-6s)$}
                                &    7444 &   7606  &   -2.18\% &  7531&  -1.16\%    \\
\multicolumn{2}{l}{$\Delta(5d_{5/2}-6p_{3/2})$}
                                &     800 &    985  &     -23\% &   759&   5.16\%     \\
  \end{tabular}
\end{ruledtabular}
\end{table}

First, we compare the In energy levels calculated using the CC and CI+all-order methods with
experiment~\cite{RalKraRea11}. The CC values include the part of the third-order energy not included
by the LCCSD method and Breit interaction in second order. The CI+all-order method
includes Breit interaction on the same footing as the Coulomb interaction, which would include some higher-order Breit
corrections. The Breit correction is small; however, it significantly improves the accuracy of the $5s5p^2~ ^4P_J$ triplet
splitting. Both calculations are carried out with $l_{max}=6$ partial waves in all intermediate sums in the
many-body expressions and include extrapolation for the contributions of the partial waves with $l>6$.

In the CC calculations, extrapolation is carried out in second order using a separate code. This second-order
calculation shows that the total contribution of the $l>6$ partial waves is close to the contribution of the single
$l=6$ partial wave. This empirical observation is used to estimate the contribution of the higher-order partial waves
in the CI+all-order approach. We find that while both methods give energy levels in very good agreement with
experiment, the CI+all-order results systematically agree better with the experimental values.

\section{Polarizabilities}
Static polarizability of the state with total angular momentum $J$, its
projection $M$, and the energy $E$ can be expressed as a sum over unperturbed intermediate states:
\begin{equation}
\alpha (J,M) = 2 \sum_n
\frac{|\langle J,M|D_z| J_n,M \rangle|^2}{E_n-E},
\label{a}
\end{equation}
where $J_n$ and $E_n$ are the total angular momenta and the energies
of the intermediate states.

The static polarizability $\alpha (J,M)$ can be conveniently decomposed
into scalar and tensor parts according to
\begin{eqnarray}
\alpha = \alpha_0 + \alpha_2 \frac{3M^2-J(J+1)}{J(2J-1)}
\label{a1}
\end{eqnarray}
where the scalar ($\alpha_0$) and tensor ($\alpha_2$)  polarizabilities
can be expressed as~\cite{KozPor99a}
\begin{eqnarray}
\alpha_0 = \frac{2}{3(2J+1)}
\sum_{n} \frac{|\langle J ||D|| J_n \rangle|^2} {E_{n}-E}.
\end{eqnarray}
and
\begin{eqnarray}
\alpha_2 &=&
4 \left (\frac{5J(2J-1)}{6(2J+3)(2J+1)(J+1)}\right)^{1/2} \nonumber \\
        &\times&\sum_{n} (-1)^{J+J_n}
        \left\{ \begin{array}{ccc} J & 1 & J_n \\ 1 & J & 2
        \end{array} \right\}
        \frac{|\langle J||D|| J_n\rangle|^2} {E_n-E},
\label{a4}
\end{eqnarray}

For an open-shell atom, $\alpha_0$ may be separated into a contribution from the valence electrons, $\alpha^v_0$,
contribution from the core electrons, $\alpha^{\rm c}$, and a core modification due to the presence of the valence
electrons, $\alpha^{vc}$. We calculate core and $\alpha^{vc}$ terms using RPA in both approaches. The valence parts are
calculated differently in the CC and CI+all-order methods.

In the CC approach, the valence polarizability of the single-electron valence state $|w\rangle$
is calculated using the sum-over-states method:
\begin{equation}
\alpha^v_0=\frac{2}{3(2j_w+1)}\sum_k\frac{{\left\langle k\left\|D\right\|w \right\rangle}^2}{E_k-E_w},
\label{eq11}
\end{equation}
where $j_w$ is the total angular momentum of the state $|w\rangle$,
${\left\langle k\left\|D\right\|w\right\rangle}$ are the single-electron reduced electric-dipole matrix elements
and sum over $k$ runs over all intermediate excited states allowed by the electric-dipole transition rules~\cite{MitSafCla10}.

In the CI+all-order approach, the valence part of the polarizability is determined  by solving the inhomogeneous
equation of perturbation theory in the valence space, which is approximated as~\cite{PorRakKoz99a}
\begin{equation}
(E - H_{\textrm{eff}})|\gamma,M'\rangle = (D_{\rm eff})_q |\gamma,J,M\rangle ,
\end{equation}
where $\gamma$ incorporates all other quantum numbers except $J$ and $M$.
The wave function $|\gamma,M' \rangle$, where $M'=M+q$, is composed of parts that have angular momenta
of $J'= J,J \pm 1$.
The construction of the effective Hamiltonian $H_\textrm{eff}$ using the all-order approach is described
in \cite{SafKozJoh09}. The effective dipole operator $D_{\textrm{eff}}$ includes RPA corrections.

\begin{table}
\caption{Contributions to the $6s$ and $5p_{1/2}$ static polarizabilities are given in $a_0^3$ in columns labeled
``$\alpha_0$'' . The experimental energies (in cm$^{-1})$ and the theoretical electric-dipole reduced matrix elements
(in a.u.) used to calculate dominant contributions are listed in columns labeled ``$\Delta E$'' and ``$D$''. The CC and
CI+all-order electric-dipole matrix elements and the polarizability contributions are listed in columns labeled ``CC''
and ``CI+All'', respectively.}
\label{tab2}
\begin{ruledtabular}
\begin{tabular}{lcrrrr}
\multicolumn{1}{l}{Contribution} &  \multicolumn{1}{c}{$\Delta E$} & \multicolumn{2}{c}{$D$} &
\multicolumn{2}{c}{$\alpha_0$} \\
 \multicolumn{1}{c}{} & \multicolumn{1}{c}{Expt.}&  \multicolumn{1}{c}{CC} &\multicolumn{1}{c}{CI+All}&
  \multicolumn{1}{c}{CC} &\multicolumn{1}{c}{CI+All} \\
 \hline
          \multicolumn{6}{c}{$6s$ polarizability}  \\
          \hline
$5p_{1/2}$     & -24373&  1.911  & 1.885& -11.0(6)&-10.7  \\
$6p_{1/2}$     & 7444  &  6.110  & 6.081& 367(12) &364   \\
$7p_{1/2}$     & 14488 &  0.683  & 0.648&  2.4(4) &2.1  \\
$8p_{1/2}$     & 17454 &  0.277  & 0.265&  0.3(1) &0.3   \\
$(9-12)p_{1/2}$&       &         &      &  0.17(3)&     \\
$(n>12)p_{1/2}$&       &         &      &  0.14(5)&      \\ [0.3pc]
$5p_{3/2}$     & -22160&  2.935  & 2.899& -28(1)  &-28  \\
$6p_{3/2}$     & 7742  &  8.529  & 8.491& 687(24) &681  \\
$7p_{3/2}$     & 14600 &  1.131  & 1.084& 6.4(1.0)&5.9   \\
$8p_{3/2}$     & 17508 &  0.489  & 0.479& 1.0(2)  &1.0        \\
$(9-12)p_{3/2}$&       &         &      &  0.6(1) &      \\
$(n>12)p_{1/2}$&       &         &      &  0.6(2) &      \\ [0.3pc]
Other          &       &         &      &         & 23\\
Core           &       &         &      & 29.6(4.0)  & 3.2 \\
VC             &       &         &      &  0.0    & 0.0    \\ [0.3pc]
Main           &       &         &      &  1024.9& 1015.5         \\
Remainder          &       &         &      &  31.1 & 26.2  \\
Total          &       &         &      & 1056(27)& 1042  \\
 \hline
\multicolumn{6}{c}{$5p_{1/2}$ polarizability}\\
\hline
 $6s           $& 24373& 1.911 & 1.885& 11.0(6)  &10.7 \\
 $7s           $& 36302& 0.548 & 0.534& 0.61(3)  &0.57 \\
 $8s           $& 40637& 0.297 & 0.297& 0.16(1)  &0.16 \\
 $(9-12)s      $&      &       &      & 0.13(3)  & \\
 $(n>12)s      $&      &       &      & 0.7(2)   & \\ [0.3pc]
 $5d_{3/2}     $& 32892& 2.623 & 2.577& 15.3(1.5)& 14.8\\
 $6d_{3/2}     $& 39049& 1.001 & 0.865& 1.9(2)  & 1.40\\
 $7d_{3/2}     $& 41836& 0.537 & 0.308& 0.5(1)   & 0.17\\
 $(8-12)d_{3/2}$&      &       &      & 0.6(2)   & \\
 $(n>12)d_{3/2}$&      &       &      & 6.1(4.0) & \\
 Other          &      &       &      &          & 31.6\\[0.3pc]
 Core\footnotemark[1] &      &       && 29.6(2.5)&  3.22\\
 VC             &      &       &      & -5.0  & -0.15\\    [0.3pc]
Main            &      &       &      &  29.4   &  27.7 \\
Remainder           &      &       &      &  32.1   &  34.6\\
 Total          &      &       &      & 61.5(5.6)&  62.4\\
\end{tabular}
\end{ruledtabular}
\small \footnotemark[1]{The uncertainty is the sum of the core and vc uncertainties}.
\end{table}

While we do not use the sum-over-state approach in the calculations of the polarizabilities in the CI+all-order method,
it is useful to calculate several dominant contributions to polarizabilities
by combining the CI+all-order values of the E1 matrix elements and energies according to the sum-over-states
formula (\ref{eq11}) above. This allows us to compare dominant terms and total remainders calculated by the CC and CI+all-order
methods.

We compare the contributions to the $6s$ and $5p_{1/2}$ polarizabilities $\alpha_0$ calculated by two approaches in
Table~\ref{tab2}. The CC and CI+all-order reduced electric-dipole matrix elements and the contributions to the
polarizability $\alpha_0$ are listed in columns labeled ``CC'' and ``CI+All'', respectively.
The experimental energies listed in column $\Delta E$ are used to
calculate the dominant contributions for consistency and to improve accuracy. We find generally very good agreement of the
CC and CI+all-order results, with the exception of the $5p_{1/2}-6d_{3/2}$ and
 the $5p_{1/2}-7d_{3/2}$ cases. These matrix elements are small and have very large correlation corrections.
The uncertainties in the CC contributions are evaluated using the well-defined approach descried in detail in
~\cite{SafSaf11}. It involves calculation of the spread of four different CC calculations of increasing accuracy for
each matrix element.  The results labeled ``Other'' in the ``CI+All'' column are obtained by subtracting the separately listed dominant
terms from the final valence value.
\begin{table}
\caption{Contributions to the $6p_{1/2}$ static polarizability are listed (in $a_0^3$) in columns labeled
``$\alpha_0$'' . The experimental energies (in cm$^{-1}$) and the theoretical electric-dipole reduced matrix elements
(in a.u.) used to calculate dominant contributions are listed in columns labeled ``$\Delta E$'' and ``$D$''. The CC and
CI+all-order matrix elements and the polarizability contributions are listed in columns labeled ``CC'' and ``CI+All'',
respectively. The contributions from the terms $6s$, $7s$, $8s$, $5d_{3/2}$, $6d_{3/2}$, and $7d_{3/2}$ are grouped
together in row ``Main''.} \label{tab6p1}
\begin{ruledtabular}
\begin{tabular}{lcrrrr}
\multicolumn{1}{l}{Contribution} &  \multicolumn{1}{c}{$\Delta E$} & \multicolumn{2}{c}{$D$} &
\multicolumn{2}{c}{$\alpha_0$} \\
 \multicolumn{1}{c}{} & \multicolumn{1}{c}{Expt.}&  \multicolumn{1}{c}{CC} &\multicolumn{1}{c}{CI+All}&
  \multicolumn{1}{c}{CC} &\multicolumn{1}{c}{CI+All} \\
 \hline
$6s$           & -7444 & 6.110 & 6.081 & -367(12)  & -363\\
$7s$           &  4485 & 6.289 & 6.239 & 645(5)    & 635\\
$8s$           &  8820 & 1.294 & 1.317 & 14        & 14\\
$(9-12)s$      &       &       &       & 5(1)      & \\
$(n>12)s$      &       &       &       & 4(1)      & \\ [0.3pc]
$5d_{3/2}$     & 1075  &10.095 & 9.893 & 6933(140) & 6659\\
$6d_{3/2}$     & 7232  &6.470  & 6.477 & 423(65)   & 424\\
$7d_{3/2}$     & 10019 &2.861  & 2.848 & 60(7)     & 59\\
$(8-12)d_{3/2}$&       &       &       & 35(11)    & \\
$(n>12)d_{3/2}$&       &       &       & 35(6)     & \\ [0.3pc]
Other          &       &       &       &           & 81\\
Core           &       &       &       & 30(4)     & 3.2\\
Main           &       &       &       & 7709(154) & 7429        \\
Remainder      &       &       &       & 108(13)   & 84 \\
Total          &       &       &       & 7817(155) &7513 \\
Recommended    &       &       &       & 7817(300)&\\
 \end{tabular}
\end{ruledtabular}
\end{table}

\begin{table*}
\caption{Contributions to the $6p_{3/2}$ scalar and tensor polarizabilities are listed (in $a_0^3$) in columns labeled
``$\alpha_0$'' and ``$\alpha_2$''. The experimental energies (in cm$^{-1}$) and the theoretical electric-dipole reduced
matrix elements (in a.u.) used to calculate dominant contributions are listed in columns labeled ``$\Delta E$'' and
``$D$''. The CC and CI+all-order matrix elements and the scalar and tensor polarizability contributions are listed in
columns labeled ``CC'' and ``CI+All'', respectively. The contributions from the terms $6s$, $7s$, $8s$, $5d_j$, $6d_j$,
and $7d_j$ are grouped together in row ``Main''.} \label{tab6p2}
\begin{ruledtabular}
\begin{tabular}{lcrrrrrr}
\multicolumn{1}{l}{Contribution} &  \multicolumn{1}{c}{$\Delta E$} & \multicolumn{2}{c}{$D$} &
\multicolumn{2}{c}{$\alpha_0$} &\multicolumn{2}{c}{$\alpha_2$} \\
 \multicolumn{1}{c}{} & \multicolumn{1}{c}{Expt.}&  \multicolumn{1}{c}{CC} &\multicolumn{1}{c}{CI+All}&
  \multicolumn{1}{c}{CC} &\multicolumn{1}{c}{CI+All} & \multicolumn{1}{c}{CC} &\multicolumn{1}{c}{CI+All}\\
 \hline
$6s$               & -7742 & 8.529 & 8.491 &-344(12)   &-341 &   344(12)& 341\\
$7s$               & 4187  & 9.413 & 9.338 & 774(7)    & 762 &  -774(7) & -762\\
$8s$               & 8522  & 1.797 & 1.830 &  14       &  14 &  -14     & -14\\
$(9-12)s$          &       &       &       &   5       &     &   -5     & \\
$(n>12)s$          &       &       &       &   4(1)    &     &   -4(1)  & \\ [0.3pc]
$5d_{3/2}$         & 777   & 4.511 & 4.420 & 958(19)   &920  &  767(15) & 736\\
$6d_{3/2}$         & 6933  & 3.157 & 3.124 &  53(7)    & 51  &   42(6)  & 41\\
$7d_{3/2}$         & 9721  & 1.350 & 1.309 &   7(1)    &  6  &    5(1)  & 5\\
$(8-12)d_{3/2}$    &       &       &       &   4(1)    &     &    3(1)  & \\
$(n>12)d_{3/2}$    &       &       &       &   4(1)    &     &    3(1)  & \\ [0.3pc]
$5d_{5/2}$         & 800   & 13.577& 13.296&8426(170)  &8080 & -1685(34)& -1616\\
$6d_{5/2}$         & 6983  & 9.218 & 9.031 &445(58)    & 427 &   -89(12)& -85\\
$7d_{5/2}$         & 9747  & 4.003 & 3.930 &  60(7)    & 58  &   -12(1) & -12\\
$(8-12)d_{5/2}$    &       &       &       &  35(7)    &     &   -7(1)  & \\
$(n>12)d_{5/2}$    &       &       &       &  33(9)    &     &   -7(1)  & \\ [0.3pc] $5s5p^2~ ^4\!P_{5/2}$
                   & 5337  &       & 2.337 &           & 37  &          & -7 \\
Other              &       &       &       &           & 79  &          & -13    \\
Core               &       &       &       &  30(4)    & 3   &          &       \\
Main               &       &       &       & 10393(180)&9979 & -1416(42)& -1367    \\
Remainder          &       &       &       &  114(12)  & 119 & -16(2)   & -20    \\
Total              &       &       &       & 10506(180)&10098 & -1432(42)&-1387       \\
Recommended        &       &       &       & 10500(400)&     & -1432(45) &  \\
 \end{tabular}
\end{ruledtabular}
\end{table*}
We note that the core contribution in the CC approach is substantially larger then the one in the CI+all-order approach
since the $5s$ shell is included in the core in the CC method, but not in the CI+all-order one. The uncertainty of the
CC core+vc term is evaluated as the difference of the DHF an RPA total core+vc values. The uncertainties in the tail
are determined based on the difference of the RPA and all-order values for terms with $n=9-12$.

As an additional comparison between two approaches, we group eight contributions to the $6s$ polarizability
($6s-(5-8)p_{1/2,3/2}$) together and list them in row ``Main''. The contributions from six terms ($6s$, $7s$, $8s$, $5d_{3/2}$,
$6d_{3/2}$, and $7d_{3/2}$) are grouped together in row ``Main'' for the $5p_{1/2}$ polarizability. These main terms
are subtracted from the totals to obtain the contributions from remaining terms. These terms are listed in
rows labeled ``Remainder''.  The difference of the remainder part for the $6s$ polarizability appears to indicate lower
value of the core polarizability when $5s^2$ is included in the core.  Overall, there is good agreement of both main
and remainder part between the two calculations. The final values for the $6s$ and $5p_{1/2}$ polarizabilities,
the $6s - 5p_{1/2}$ Stark shift, and their uncertainties are discussed in the next section.

Contributions to the $6p_{1/2}$ and $6p_{3/2}$ polarizabilities of In (in $a_0^3$) are listed in Tables~\ref{tab6p1} and
\ref{tab6p2}, respectively. The tensor  polarizability $\alpha_2$ of the $6p_{3/2}$ state is given in
Table~\ref{tab6p2}. These tables are structured in exactly the same way as Table~\ref{tab2}. The results of both CC and
CI+all-order calculations are given. The only difference is the listing of the $5s5p^2 ~^4\!P_{5/2}$  contribution  to
the $6p_{3/2}$ polarizability. The contributions of the $5s5p^2~ ^4\!P_{J}$ state to all other polarizabilities
considered here were found to be negligible.

We find 3.9\% difference between the CC and CI+all-order values for both scalar
$6p_j$ polarizabilities and 3.1\% difference between the values of the tensor $\alpha_2$ polarizability for the
$6p_{3/2}$ state. These differences are caused by the 2\% difference in the values of the $6p_{1/2}-5d_{3/2}$,
$6p_{3/2}-5d_{3/2}$, and $6p_{3/2}-5d_{5/2}$ matrix elements which  dominate the $6p_j$ polarizability values. This 2\%
difference is consistent with the expected accuracy of the CI+all-order method for these transitions.

We evaluate the
accuracy of the CC values to be on the order of 1\%. The uncertainty evaluation is carried out differently for these
three transitions owing to a convergence problem of the all-order equations for the $5d$ states. We performed the
calculations with three and five iterations in the LCCSD approximation and in the LCCSDpT approximation,
and carried out the scaling procedure (described, for example, in Ref.~\cite{SafJoh08})
using these four different starting points. The spread
of the resulting scaled values is 0.7\%. Since the scaling estimates the dominant omitted corrections in such
transitions, all other omitted corrections should not exceed this upper bound of 0.7\%, resulting in total uncertainty
estimate of 1\%. The CC and CI+all-order values for all other contributions, including small remainders, were found to be in
a good agreement. Because we expect the CC method to yield more accurate values of the $6s-5d$ matrix elements, we take
the CC values of the $6p$ polarizabilities as the final results.

Most likely, the discrepancy with the CI+all-order calculations is
caused by an omission of the small corrections to the effective dipole operator in the CI+all-order approach.
However, it might be possible that the $6s-5d$ matrix elements are affected by the small mixing of the even $5s^2 5d$ states with
the $5s5p^2~^4\!P_J$ triplet, which is accounted for by the CI+all-order method, but not the CC method. The weight
(in probability) of the $5s5p^2$ configuration in the $5d_j$ levels is $0.02-0.03$, i.e., small but non negligible.
Moreover, there is a very strong mixing
of the $nd$ configurations.  Therefore, we take the difference of the CI+all-order and CC results as the final uncertainty to
account for the possible uncertainty due to such mixing. We note that $5s^2 6s$ state  essentially does not mix with
the $5s5p^2$ configuration (its weight is only 0.0003), so $6s-6p$ matrix elements are not affected by such mixing.
A high-precision measurement of the $6p_{3/2}-6s$ Stark shift should resolve this question. The final values for the
$6p_{3/2}$ polarizabilities are listed in the last rows of Tables~\ref{tab6p1} and \ref{tab6p2}.
\begin{table*}
\caption{Final values of the $6s$ and $5p_{1/2}$ polarizabilities and their difference $\Delta \alpha_0$(a.u.).
Determination of the reduced electric-dipole $6s-6p_j$ matrix elements (in a.u.) and $6p_j$ lifetimes (in ns) from the
combination of the recently measured Stark shift ~\cite{RanSchLor13} and theoretical values. The quantity $C$ is the
value of $\Delta \alpha_0(6s-5p_{1/2})$ with the contribution of $6s-6p_j$ transitions subtracted out. The results are
compared with other theory and experiment.} \label{tab3}
\begin{ruledtabular}
\begin{tabular}{lcccccccc}
\multicolumn{1}{l}{} &  \multicolumn{1}{c}{$\alpha_0(6s)$} & \multicolumn{1}{c}{$\alpha_0(5p_{1/2})$} &
\multicolumn{1}{c}{$\Delta \alpha_0(6s-5p_{1/2})$} & \multicolumn{1}{c}{$C$} & \multicolumn{1}{c}{$D(6s-6p_{1/2})$}&
\multicolumn{1}{c}{$D(6s-6p_{3/2})$} &\multicolumn{1}{c}{$\tau(6p_{1/2})$}&
\multicolumn{1}{c}{$\tau(6p_{3/2})$}  \\
\hline
CC                 &1056     &61.5     & 995        & -59.6& 6.126    & 8.551    &          &          \\
CI+All             &1042     &62.4     & 980        & -65.2& 6.141    & 8.575    &          &           \\
Final              &1056(20) &61.5(1.3)& 995(21)    & -59.6(7.8)& 6.126(24)& 8.551(34)& 63.77(50)& 58.17(45)\\
Expt.~\cite{RanSchLor13}  &         &         & 1000.2(2.7)&      &          &          &          &           \\
Expt.~\cite{GueMilBed84}   &         & 69(8) &              &      &          &          &          &           \\
Theory~\cite{Fle05}&         & 61.48 &              &      &          &          &            &         \\
Theory~\cite{BorZelEli12}&         & 62.0(1.9)&            &      &          &          &          &             \\
Theory~\cite{SahDas11}  &         &        &             &       &         &           & 63.8(8)  &  58(1)   \\
\end{tabular}
\end{ruledtabular}
\end{table*}

\section{Discussion of the results and their uncertainties}

While we have estimated the uncertainties of the CC calculation, it is possible to improve our evaluation of the
uncertainties by comparing the CC and CI+all-order results. As we have described above, these two high-precision
approaches include somewhat different higher-order effects. The CI+all-order calculations include the valence-valence
correlation corrections to the wave functions very precisely, as indicated by excellent agreement of the respective
energies with experiment. On the other hand, the effective dipole operator $D_{\rm eff}$
includes only RPA corrections in the CI+all-order method at the present time, omitting the structure radiation,
normalization, and other small corrections.
These corrections to the electric-dipole matrix elements are included in the CC method.

In the polarizability calculations, we use the experimental energies for the main terms. Therefore, more accurate
method of calculating individual matrix elements is somewhat more important for the $6s$ polarizability.
In the framework of the CC method, the $5s$ shell is included in the core and we have the large core polarizability. It
leads to increasing the total uncertainty of the polarizability because the core polarizability is calculated with less
accuracy. But it cancels out when the Stark shift of a transition is calculated. The CI+all-order method treats
contributions with high $n$ with better accuracy, since it is done by solving the inhomogeneous equation instead of
using the sum-over-states method. This is not significant for the $6s$ polarizability where such tail contributions are
small, but is important for the $5p_{1/2}$ polarizability.

In summary, the CC and CI+all-order methods together include all correlation corrections that are expected to be
dominant for the present polarizability calculations. Therefore, the uncertainty can be approximated as the difference
$\delta \alpha=|\alpha\textrm{(CC)}-\alpha\textrm{(CI+all)}|$ of the CC and CI+all-order results. All other omitted
higher-order corrections are expected to be smaller than already included ones, therefore, we can also assume that they
do not exceed $\delta \alpha$. Therefore, we take our final uncertainty in the polarizabilities and their difference to
be $\sqrt{2}\delta\alpha$, calculated separately for each of the properties. We list our final values of the $6s$ and
$5p_{1/2}$ polarizabilities and their difference $\Delta \alpha_0(6s-5p_{1/2})$ in Table~\ref{tab3} in a.u.

We take the CC results to be the final values since CC method accounts for more higher-order corrections to the E1
matrix elements that dominate the $6s$ polarizability.  For consistency, CC value of the $5p_{1/2}$ polarizability is
used when calculating the final value for the $6s-5p_{1/2}$ Stark shift. Our final result is in excellent agreement
with recent high-precision measurement of the $6s-5p_{1/2}$ Stark shift ~\cite{RanSchLor13}, which allows for benchmark
comparison of the theoretical approachers. We find that the CC value is closer to the experimental measurement than the
CI+all-order result. Our calculated polarizability of the $5p_{1/2}$ state is in excellent agreement with recent
coupled-cluster single, double, and perturbative tripes excitation [CCSD(T)] calculations of
Refs.~\cite{BorZelEli12,Fle05}. We note that our implementation of the coupled-cluster method differs significantly
from that of Refs.~\cite{BorZelEli12,Fle05} (see recent review~\cite{MitSafCla10} for details).

\section{Determination of the lifetimes}

Recent precision measurement of the $6s-5p_{1/2}$ Stark shift ~\cite{RanSchLor13} can be combined with the present
calculations to obtain very accurate lifetimes of the $6p_{1/2}$ and $6p_{3/2}$ states. This is possible since the
$6s-6p_{j}$ matrix elements overwhelmingly dominate the values of this Stark shift as illustrated by Table~\ref{tab2}.
Separating the $6s-6p_j$ contributions (see Eq.(\ref{eq11})), we write the $\Delta \alpha_0(6s-5p_{1/2})$ Stark shift
as
\begin{equation}
\label{eq14} \Delta \alpha_0(6s-5p_{1/2})=B S+C,
\end{equation}
where
\begin{equation}
B=\frac{1}{3}\left( \frac{1}{E(6p_{1/2})-E(6s)} + \frac{R^2}{E(6p_{3/2})-E(6s)} \right),
\label{B}
\end{equation}
$S=D^2$ is the $6s-6p_{1/2}$ line strength, $R$ is the ratio of the $D(6s-6p_{3/2})$ and $D(6s-6p_{1/2})$ reduced E1
matrix elements, and the term  $C$ contains all other contributions to the Stark shift. We calculate the ratio $R$ to
be 1.396(1). The uncertainty (0.001) is very  small since the ratio $R$ is very insensitive to different corrections.
Using the results presented in Table~\ref{tab2}, we see that the CC and CI+all-order methods give $R$ equal to each
other up to fourth significant figure. Substituting the ratio $R$ and the corresponding experimental energies in
Eq.~(\ref{B}) gives $B=28.24(3)$ a.u.

The values of $C$ in the CC and CI+all-order methods are obtained from the results presented in Table~\ref{tab2} and
are equal to -59.6 a.u. and -65.2 a.u., respectively (see Table~\ref{tab3}). Adding the  relevant uncertainties from CC
calculations of Table~\ref{tab2} in quadrature leads to the uncertainty $\delta C= 4.9$ a.u. This number is consistent
with the difference of the CC and CI+all-order values, 5.6 a.u. We note that the uncertainty of the core term does not
contribute to $\delta C$, since this term is canceled when the Stark shift is calculated. The $5p_{1/2}$ $\alpha_{vc}$
term does contribute, and its uncertainty is 1.3~a.u.
To evaluate the final uncertainty in $C$ we use the same rule  as for the polarizability: multiply the difference of
the CC and CI+all-order values by $\sqrt{2}$ to account for other small uncertainties not included in our
consideration. Again, we assume that they cannot exceed the difference of the CC and CI+all-order values. The final
value for $C$ is presented in Table~\ref{tab3}.

 There are three sources of the uncertainties contributing to the
uncertainty in $D(6s-6p_{1/2})$: uncertainties in $C$, $B$, and experimental values of $\Delta \alpha_0$.  For
convenience, we calculate first the uncertainty in line strength $S$ using formula:
\begin{equation}
\delta S=\frac{1}{B}\sqrt{(\delta C)^2+(\delta \Delta\alpha_0)^2+(S \delta B)^2}.
\end{equation}
The relative uncertainty in $D$ is a half of the relative uncertainty in $S$. The uncertainty in $B$ is negligible. The
final values of the matrix elements are listed in Table~\ref{tab3}. The lifetimes of the $6p_{1/2}$ and $6p_{3/2}$
states are obtained using the formula $\tau_a=1/A_{ab}$ since there is only one $E1$ decay channel for each state. The
transition rate $A_{ab}$ is given by
\begin{equation}
A_{ab}=\frac{2.02613\times10^{18}}{\lambda_{ab}^3}\frac{S_{ab}}{2J_a+1}\,\,\text{s}^{-1},
\end{equation}
where the transition wavelength $\lambda_{ab}$ is in \AA~.
The relative uncertainties
in the lifetimes are twice that of the relative uncertainties of the corresponding E1 matrix elements.  The final
values are given in Table~\ref{tab3}.

To simplify the extraction of the $5d_j$ lifetimes from future experimental value of the $6p_{3/2}-6s$ Stark shift, we
evaluated the sum of all contributions to the $6p_{1/2}$ and $6p_{3/2}$ polarizabilities with the $5d-6p$ terms
excluded. These quantities, which we designate as $\widetilde{C}(6p_j)$, are obtained from the data in
Tables~\ref{tab6p1} and \ref{tab6p2}. We note that $\widetilde{C}(6p_{1/2})$ and $\widetilde{C}(6p_{3/2})$ refer to the
contributions to the polarizabilities, rather than their differences as in Eq.~(\ref{eq14}).

We find that the CC and CI+all-order results are very close together, and are well within the uncertainty estimates of
the CC data. The CC values (in a.u.) are $\widetilde{C}(6p_{1/2})=884(68)$, $\widetilde{C}_0(6p_{3/2})=1123(60)$,  and
$\widetilde{C}_2(6p_{3/2})=-514(19)$, where $S_0$ and $S_2$ are related to the scalar and tensor polarizabilities,
respectively. The corresponding CI+all-order values are 854, 1098 and  $-$506 (in a.u.). The ratio of the
$6p_{1/2}-5d_{3/2}$ and $6p_{3/2}-5d_{3/2}$ matrix elements is $2.238(4)$ and the ratio of the $6p_{3/2}-5d_{5/2}$ and
$6p_{3/2}-5d_{3/2}$ matrix elements is 3.0095(16).

\section{Conclusion}

We carried out a first systematic comparison of the linearized coupled-cluster and CI+all-order method using the
polarizabilities of the low-lying states of the In atom as a benchmark testing case. We find that the CI+all-order
method produces more accurate data for the energy levels. It appears that the CC data for the E1 matrix elements are
somewhat more accurate due to more complete inclusion of the small higher-order corrections to the matrix elements in
the cases where relevant configuration mixing of trivalent states is negligible. This is an additional motivation to
incorporate such corrections into the CI+all-order formalism at the all-order level in the future.

Our result for the $6s-5p_{1/2}$ Stark shift is in excellent agreement with the recent high-precision experiment
\cite{RanSchLor13}. We also provide predictions for the polarizabilities of the $6p_{1/2}$ and $6p_{3/2}$ states. A
precise experimental measurement of the $6p_{3/2}-6s$ Stark shift proposed in \cite{RanSchLor13} would be a good test
of our calculations. This will also provide an excellent test of the theoretical approaches. Combining the present
calculations with the experimental Stark shift data allows very accurate extraction for the lifetimes of the low-lying
In states.

\section*{Acknowledgements}
We thank P.~K.~Majumder for bringing this problem to our attention and helpful discussions. We thank M.~G. Kozlov for
discussion of the CI+all-order calculations and construction of the CI spaces and A.~Borschevsky for providing
unpublished CCSD and CCSD(T) values for the In$^+$ core polarizabilities.  The work of M.S.S was supported in part by
the NSF Grant No. PHY-1068699. The work of S.G.P. was supported in part by US NSF Grant No.\ PHY-1068699 and RFBR Grant
No. 11-02-00943.

 \end{document}